\documentclass[12pt,twoside]{article}
\usepackage{amsmath}
\usepackage{graphicx}
\usepackage{caption,subcaption}
\usepackage{epsfig}
\usepackage{epstopdf}
\usepackage{xcolor}

\numberwithin{equation}{section}
\begin{document}
	\setcounter{page}{1}
	%\newcounter{equation}[section]
	\newtheorem{t1}{Theorem}[section]
	\newtheorem{d1}{Definition}[section]
	\newtheorem{c1}{Corollary}[section]
	\newtheorem{l1}{Lemma}[section]
	\newtheorem{r1}{Remark}[section]

	\newcommand{\cA}{{\cal A}}
	\newcommand{\cB}{{\cal B}}
	\newcommand{\cC}{{\cal C}}
	\newcommand{\cD}{{\cal D}}
	\newcommand{\cE}{{\cal E}}
	\newcommand{\cF}{{\cal F}}
	\newcommand{\cG}{{\cal G}}
	\newcommand{\cH}{{\cal H}}
	\newcommand{\cI}{{\cal I}}
	\newcommand{\cJ}{{\cal J}}
	\newcommand{\cK}{{\cal K}}
	\newcommand{\cL}{{\cal L}}
	\newcommand{\cM}{{\cal M}}
	\newcommand{\cN}{{\cal N}}
	\newcommand{\cO}{{\cal O}}
	\newcommand{\cP}{{\cal P}}
	\newcommand{\cQ}{{\cal Q}}
	\newcommand{\cR}{{\cal R}}
	\newcommand{\cS}{{\cal S}}
	\newcommand{\cT}{{\cal T}}
	\newcommand{\cU}{{\cal U}}
	\newcommand{\cV}{{\cal V}}
	\newcommand{\cX}{{\cal X}}
	\newcommand{\cW}{{\cal W}}
	\newcommand{\cY}{{\cal Y}}
	\newcommand{\cZ}{{\cal Z}}
	
	\newcommand{\mathbb}[1]{\mbox{\boldmath $#1$}}
	\def\WW{\begin{stack}{\circle \\ W}\end{stack}}
	\def\ww{\begin{stack}{\circle \\ w}\end{stack}}
	\def\st{\stackrel}
	\def\Ra{\Rightarrow}
	\def\R{{\mathbb R}}
	\def\bi{\begin{itemize}}
		\def\ei{\end{itemize}}
	\def\bt{\begin{tabular}}
		\def\et{\end{tabular}}
	\def\lf{\leftarrow}
	\baselineskip 20truept
	\begin{center}
		{\Large \bf  Shared Frailty Models Based on Cancer Data}\\
		\bigskip
		{\bf Shikhar Tyagi$^{a}$, Arvind Pandey$^{a}$, David D. Hanagal$^{b}$\footnote{Corresponding author e-mail: david.hanagal@gmail.com }.     \\
	\small	$^{a}$Department of Statistics, Central University of Rajasthan, Rajasthan, India\\
	\small $^{b}$ Department of Statistics, Savitribai Phule Pune University, Pune, India.}	\\	
			\bigskip

	\end{center}
\begin{abstract}
	
Traditional survival analysis techniques focus on the occurrence of failures over the time. During analysis of such events, ignoring the related unobserved covariates or heterogeneity involved in data sample may leads us to adverse consequences. In this context, frailty models are the viable choice to investigate the effect of the unobserved covariates. In this article, we assume that frailty acts multiplicatively to hazard rate. We propose inverse Gaussian (IG) and generalized Lindley (GL) shared frailty models with generalized Weibull (GW) as baseline distribution in order to analyze the unobserved heterogeneity. To estimate the parameters in models, Bayesian paradigm of Markov Chain Monte Carlo technique has been proposed. Model selection criteria have been used for the comparison of models. Three different cancer data sets have been analyzed using the shared frailty models. Better models have been suggested for the data sets.
	
\end{abstract}


\section{\large Introduction}
In survival data, a common approach is that each individual under study experiencing the same risk factors which act as multiplicatively. Cox (1972) proposed proportional hazard model with the simplest case of independent and identically distributed data and is based on the assumption that the study population is homogeneous. Sometimes, in real-life situations risk (hazard rate) changes from one family to another family, one group to another group, one cluster to another cluster. In general, individuals differ greatly because of genetic factor and environmental factors. This heterogeneity is often referred to as variability and it is generally recognized as one of the most important sources of variability in medical and biological applications. Heterogeneity in the population exists, and this heterogeneity is called as a frailty.
The key idea of these models is that individuals have different heterogeneities (or frailties), and that the most frail will die earlier than the lesser frail. Consequently, systematic selection of robust individuals takes place, which distorts what is observed. When mortality rates are estimated, one may be interested in knowing the change in the behavior over time or age. Quite often, they rise at the beginning of the observation period, reach a maximum and then decline (unimodal intensity) or level off.

Frailty models are extensively used in the survival analysis to account for the unobserved heterogeneity in individual risks to disease and death [Vaupel et al. 1979]. The frailty model is usually modeled as an unobserved random variable acting multiplicatively on the baseline hazard function. To analyze the bivariate data on related survival times (e.g. matched pairs experiments, twin or family data), the shared  frailty models were suggested.  The frailty model is a random effect model for time to event data which is an extension of the Cox's proportional hazards model. Bivariate survival data arises whenever each study subjects experience two events. Particular examples include failure times of paired human organs, (e.g. kidneys, eyes, lungs, breasts, etc.) and the first and the second occurrences of a given disease. In the medical literature, several authors considered paired organs of an individual as a two-component system, which work under interdependency circumstances. In industrial applications, these data may come from systems whose survival depend on the survival of two similar components.

If covariates are known, they can be included in the analysis, for example by using the
proportional hazards model as described above. But it is nearly always impossible to
include all important risk factors, perhaps because we have little or no information on
the individual level (this applies for example to population studies, where often the only
known risk factors are sex and age). Furthermore, we may not know the relevance of
the risk factor or even that the factor exists. In other cases it may be impossible to
measure the risk factor without great financial cost or time effort. In such cases, two
sources of variability in duration data are useful to consider: variability accounted for by
observable risk factors (which is thus theoretically predictable) and heterogeneity caused
by unknown covariates and which is therefore theoretically unpredictable even when all
relevant information at that time is known. It is the latter which is of specific interest
here, and the subject of observable covariates is treated here only for completeness.
There are advantages in considering these two sources of
variability separately: heterogeneity explains some `unexpected' results or gives an alternative
explanation of some results, for example non-proportional or decreasing hazards.
If some individuals experience a higher risk of failure, then the remaining individuals
at risk tend to form a more or less selected group with lower risk. Ignoring frailty may have adverse consequences.
An estimate of the individual hazard rate without taking into account the unobserved frailty will thus
underestimate the hazard function to an increasingly greater extent as time goes by.

The extension of Cox (1972) proportional hazard model after incorporating unobserved covariates can be written as
\begin{eqnarray}
\phi(z|\underline{K})&=& \phi_{0}(z)e^{\underline{K}^{'}\underline{\beta_{0}}+\underline{V}^{'}\underline{\beta_{1}}}
\end{eqnarray}
where, $\phi(t|\underline{K})$ stands for conditional hazard rate, $\phi_{0}(z)$ stands for baseline hazard rate. $\underline{K}^{'}=(K_{1j},K_{2j},..., K_{mj})$, $\underline{V}^{'}=(V_{1j},V_{2j},..., V_{mj})$ are considered as the vector of known and unknown covariates respectively, $\underline{\beta_{0}}$ and $\underline{\beta_{1}}$ are indicated as the vector of corresponding regression coefficients of order m. $W=e^{\underline{V}^{'}\underline{\beta_{1}}}$ called as frailty effect.\\
A random impact that is unobservable risk shared by the subject characterized as frailty which was introduced by Vaupel et al. (1979). To handle such kind of problems, many models have been derived in survival analysis.
Clayton's (1978) random effect model of the bivariate survival was a key innovation. He introduced the notion of the shared relative risk. This model was further developed
by Oakes (1982) to analyze the association between two non-negative random variables. Hougaard (1985,1991,2000) had discussed the different aspects of frailty on a broad scale.  In the last decade, frailty regression models in mixture distribution have
been discussed by Hanagal (2008). Hanagal and Dabade (2013, 2015) proposed modelling of the
inverse Gaussian frailty model and comparison of different frailty models for analyzing kidney
infection data. Modelling kidney infection data for inverse Gaussian shared frailty was done by
Hanagal and Pandey (2014a). Hanagal and Pandey (2017a) were used the shared inverse Gaussian
frailty models based on additive hazard. For reversed hazard rate setup, Hanagal and Pandey
(2014b,2015b,2016a,2016b,2017b) have contemplated gamma and inverse Gaussian shared frailty
models with different baseline distribution functions. Hanagal and Sharma (2013, 2015a, 2015b,
2015c) analyzed acute leukemia data, kidney infection data and diabetic retinopathy data using
shared gamma and inverse Gaussian frailty models for the multiplicative model. Hanagal (2011, 2017, 2019) gave extensive literature review on
different shared frailty models. Pandey et al. (2020) investigated reversed hazard rate based generalized inverse Gaussian shared frailty models. On the basis of reversed hazard rate, Pandey et al. (2021) and Tyagi et al. (2021) built different generalized Lindley shared frailty models. Inverse Gaussian, weighted Lindley, and generalized Lindley shared frailty models were established by Pandey et al. (2020), Tyagi et al. (2021), and Pandey et al. (2021) respectively.
The main aim of this article has three objectives. First, Inverse Gaussian (IG) and generalized Lindley (GL) frailty models for multiplicative hazard rate with generalized Weibull (GW) as baseline distribution have been introduced. Second, Bayesian approach of estimation has been employed to estimate the unknown parameters under random censoring. Third, simulation study and data analysis have been done for the bladder cancer, lung cancer, and ovarian cancer data sets.
\section{Model Description and Notations}
For the study, $n$ individuals has been considered. Let a random variable $z_j$ that denotes the survival time corresponding to $j^{th}$ individuals $(j= 1, 2,..., n)$. $\underline{K}_{j} = (k_{1},..., k_{m})$ is a vector of  known covariates for $j^{th}$ individual. For the $j^{th}$ individual, $W_j$ be the frailty. For given frailty $W_j= w_j$, at survival time $z_{j}\in{\rm I\!R^{+}}$, the form of conditional hazard function corresponding to $j^{th}$ individual can be defined as,
\begin{eqnarray}
\phi\left(z_j| w_j, \underline{K}_{j} \right) &=& w_{j}\phi_{0}\left(z_{j} \right)\rho_{j}
\end{eqnarray}
here, $\rho_{j}$ stands for $e^{\underline{K}^{'}_{j}\underline{\beta}_{0}}$, corresponding cumulative hazard rate for $j^{th}$ can be obtained as,
\begin{eqnarray}
\Phi\left(z_j| w_j, \underline{K}_{j} \right) &=& w_{j}\Phi_{0}\left(z_{j} \right)\rho_{j}
\end{eqnarray}
here, cumulative baseline hazard function is symbolised by $\Phi_{0}$, corresponding conditional survival function for frailty $W_j=w_j$ is,
\begin{eqnarray}
\Lambda\left( z_j| w_j, \underline{K}_{j} \right) &=& e^{-\Phi\left(z_j| w_j, \underline{K}_{j} \right)} \notag \\
&= & e^{- w_{j}\Phi_{0}\left(z_{j} \right)\rho_{j} }
\end{eqnarray}
where,$\Lambda(.)$ stands for survival function, and suppose, $f_{W}(w_j)$ stand for frailty density function. After integrating over the frailty variable, for $j^{th}$ variable, unconditional survival function can be obtained,
\begin{eqnarray}
\Lambda(z_{j}) & = & \int_{w_j\in{\rm I\!R^{+}}}^{} \Lambda(z_j|w_j, \underline{K}_{j})f_{W}(w_j)dw_j \notag \\
& = & L_{W_j}\left( \Phi_{0}(z_j)\rho_{j} \right)
\end{eqnarray}
where, $L_{W_j}(.)$ is Laplace transformation of frailty variable $W_j$.

\section{\large Frailty Distributions}
\subsection{Inverse Gaussian Distribution}
Firstly, Hougaard (1984) introduced the inverse Gaussian (IG) distribution as an alternative of gamma frailty distribution. IG distribution is associated to exponential family, and because of that it provides unimodel density. For a early age failure situations, it provides much flexibility in modeling. Its failure rate is expected to be non-monotonic. The inverse Gaussian distribution has shape resembles the other skewed density functions, such as log-normal and gamma. Due to these properties, we encourage to employ IG as a frailty distribution. A non-negative, continuous random variable $W\in{\rm I\!R^{+}}$ is said to have a IG distribution with parameters $\varepsilon$ and $\eta$, if its density function is,
\[
f_{W}(w) = \left\{ \begin{array}{ll}
\left(\frac{1}{2 \pi \eta}\right)^{\frac{1}{2}} w^{-\frac{3}{2}}\exp\left\lbrace \frac{-(w-\varepsilon)^{2}}{2w\eta \varepsilon^{2}} \right\rbrace  & ; w\in{\rm I\!R^{+}},\varepsilon,\eta \in{\rm I\!R^{+}}\\
0 & ; \textit{otherwise}
\end{array}
\right.
\]
and the Laplace transform is,
\begin{eqnarray}
L_{W}(s)&=& \exp\left[\frac{1}{\varepsilon \eta}-\left(\frac{1}{\varepsilon^{2} \eta^{2}}+\frac{2s}{\eta} \right)^{\frac{1}{2}}  \right].
\end{eqnarray}
The mean of frailty variable is $E[W]= \varepsilon$ and corresponding variance is $V(W)=\varepsilon^{3}\eta$. To avoid problems of identifiability, we consider $E[W]=1$. Consequently, the density function, Laplace transformation and variance for IG distribution reduced to,
\[
f_{W}(w) = \left\{ \begin{array}{ll}
\left(\frac{1}{2 \pi \eta}\right)^{\frac{1}{2}} w^{-\frac{3}{2}}\exp\left\lbrace \frac{-(w-1)^{2}}{2w\eta} \right\rbrace  & ; w\in{\rm I\!R^{+}},\eta \in{\rm I\!R^{+}}\\
0 & ; \textit{otherwise}
\end{array}
\right.
\]
\begin{eqnarray}
L_{W}(s)&=& \exp\left[\frac{1-(1+2\eta s)^{\frac{1}{2}}}{\eta}  \right]
\end{eqnarray}
To get unconditional survival function, replacing the Laplace transformation in equation (3.2),
\begin{eqnarray}
\Lambda(z_{j})&=& exp\left[\frac{1-(1+2\eta \Phi_{0}(z_j)\rho_{j})^{\frac{1}{2}}}{\eta}  \right]
\end{eqnarray}
\subsection{Generalized Lindley Distribution}
The Lindley distribution with one parameter was first proposed by Lindley (1958) for analyzing failure times data. It belongs to an exponential family, but it is used as an alternative to the exponential distribution. Lindley distribution is alluring due to the ability of modeling failure time data with increasing, decreasing, unimodal and bathtub shaped hazard rates. Ghitany et al. (2008) and Ghitany et al. (2011) discussed different properties of Lindley distribution and also showed that Lindley distribution is better than the exponential distribution for modeling failure time data when considering hazard rate is unimodal or bathtub shaped. It is also shown that Lindley distribution is more flexible than exponential distribution in modeling lifetime data. The new generalized Lindley distribution can be defined as,
\begin{eqnarray}
f(w)&=& pf_{1}(w)+(1-p)f_{2}(w)
\end{eqnarray}
where, $f_{1}(w)$ and $f_{1}(w)$ are $G(\mu,\eta)$ and $G(\lambda,\varepsilon)$ distributions respectively and $p$ is the mixing proportion and are given by,
\[
f_{1}(w) = \left\{ \begin{array}{ll}
\frac{w^{\frac{1}{\eta}}e^{-\frac{w}{\mu}}}{\mu^{\frac{1}{\eta}}\Gamma\left(\frac{1}{\eta} \right) } &; w\in{\rm I\!R^{+}},\mu,\eta \in{\rm I\!R^{+}}\\
0 & ; \textit{otherwise}
\end{array}
\right.
\]

\[
f_{2}(w) = \left\{ \begin{array}{ll}
\frac{w^{\frac{1}{\varepsilon}}e^{-\frac{w}{\lambda}}}{\lambda^{\frac{1}{\varepsilon}}\Gamma\left(\frac{1}{\varepsilon} \right) }  &; w\in{\rm I\!R^{+}},\varepsilon,\lambda \in{\rm I\!R^{+}}\\
0 & ; \textit{otherwise}
\end{array}
\right.
\]
The mean, variance and Laplace transform of $W$ are given by,
 \begin{eqnarray}
 E(W)&=& p\frac{\mu}{\eta}+(1-p)\frac{\lambda}{\varepsilon}
 \end{eqnarray}
 \begin{eqnarray}
 V(W)&=& p\frac{\mu^{2}}{\eta}+(1-p)\frac{\lambda^{2}}{\varepsilon}
 \end{eqnarray}
 \begin{eqnarray}
 L_{W}(s)&=& p(1+s\mu)^{-\frac{1}{\eta}}+(1-p)(1+s\lambda)^{-\frac{1}{\varepsilon}}
 \end{eqnarray}
 To resolve identifiability problem, we assume expected value of $W$ equal to one which
 leads to $\mu=\eta$ and $\lambda=\varepsilon$. Substituting these in equation (3.4), we get
 \begin{eqnarray}
 f_{W}(w)= \left( \frac{\eta}{(\eta+\varepsilon)}\right) \frac{w^{\frac{1}{\eta}}e^{-\frac{w}{\eta}}}{\eta^{\frac{1}{\eta}}\Gamma\left(\frac{1}{\eta} \right)}+\left( \frac{\varepsilon}{(\eta+\varepsilon)}\right) \frac{w^{\frac{1}{\varepsilon}}e^{-\frac{w}{\varepsilon}}}{\varepsilon^{\frac{1}{\varepsilon}}\Gamma\left(\frac{1}{\varepsilon} \right) }
 \end{eqnarray}
 where, $p=\left( \frac{\eta}{(\eta+\varepsilon)}\right)$.
 The variance and Laplace transform of $W$ with restriction $E(W)=1$ are,
 \begin{eqnarray}
 V(W)&=& \frac{\eta^{3}+\varepsilon^{3}}{(\eta+\varepsilon)^{2}}
 \end{eqnarray}
 \begin{eqnarray}
 L_{W}(s)&=&\frac{\eta(1+s\eta)^{-\frac{1}{\eta}}+\varepsilon(1+s\varepsilon)^{-\frac{1}{\varepsilon}}}{(\eta+\varepsilon)}
 \end{eqnarray}
To get unconditional survival function, replacing the Laplace transformation in equation (3.10),
\begin{eqnarray}
\Lambda(z_{j})&=& \frac{\eta(1+\Phi_{0}(z_j)\rho_{j}\eta)^{-\frac{1}{\eta}}+\varepsilon(1+\Phi_{0}(z_j)\rho_{j}\varepsilon)^{-\frac{1}{\varepsilon}}}{(\eta+\varepsilon)}
\end{eqnarray}

\section{\large Baseline Distributions}
\subsection{Generalized Weibull Distribution}
Here, the generalized Weibull (GW) distribution has been postulated as a baseline distribution. If a continuous random variable $Z\in{\rm I\!R^{+}}$ follows the GW distribution then the survival, hazard, and cumulative hazard function, are respectively,	
\begin{eqnarray}
\Lambda(z)&=& \left\{ \begin{array}{ll}
1-\left(1- e^{-\delta z^{\xi}} \right)^{\zeta} &;z\in{\rm I\!R^{+}}, \delta, \zeta, \xi\in{\rm I\!R^{+}} \\
1 &; \textit{otherwise}
\end{array} \right.
\end{eqnarray}
\begin{eqnarray}
\phi_{0}(z)&=& \left\{ \begin{array}{ll}
\frac{\xi \zeta \delta z^{\xi-1}e^{-\delta z^{\xi} (1-e^{-\delta z^{\xi}})^{\zeta-1}}}{1-\left(1- e^{-\delta z^{\xi}} \right)^{\zeta}} &;z\in{\rm I\!R^{+}}, \delta, \zeta, \xi\in{\rm I\!R^{+}}\\
1 &; \textit{otherwise}
\end{array} \right.
\end{eqnarray}
\begin{eqnarray}
\Phi_{0}(z)&=& \left\{ \begin{array}{ll}
-\log \left(1- \left(1- e^{-\delta z^{\xi}} \right)^{\zeta} \right) &;z\in{\rm I\!R^{+}}, \delta, \zeta, \xi\in{\rm I\!R^{+}} \\
0 &; \textit{otherwise}
\end{array} \right.
\end{eqnarray}

\section{\large Proposed Frailty Models}
Cumulative hazard function of GW distribution from (4.3) has been substituted to get model for hazard function in equation (3.3) and (3.11),
\begin{eqnarray}
\Lambda_{1}(z_{j})&=& exp\left[\frac{1-(1-2\eta \log \left(1- \left(1- e^{-\delta z_{j}^{\xi}} \right)^{\zeta} \right)\rho_{j})^{\frac{1}{2}}}{\eta}  \right]
\end{eqnarray}
  {\fontsize{10}{10}\selectfont
\begin{eqnarray}
\Lambda_{2}(z_{j})&=& \frac{\eta\left( 1-\log \left(1- \left(1- e^{-\delta z_{j}^{\xi}} \right)^{\zeta} \right)\rho_{j}\eta\right) ^{-\frac{1}{\eta}}+\varepsilon\left( 1-\log \left(1- \left(1- e^{-\delta z_{j}^{\xi}} \right)^{\zeta} \right)\rho_{j}\varepsilon\right) ^{-\frac{1}{\varepsilon}}}{(\eta+\varepsilon)}
\end{eqnarray}}
here, equations (5.1)and (5.2) can be called as IG-GW Model and GL-GW Model respectively that have been established for IG and GL frailty distributions with GW baseline distribution.

\section{\large Likelihood Functions under Random Censored Data}
For the study, n individuals has been considered. Observed failure time has been indicated by $Z_j$. We are using random censoring scheme. Censoring time, supposed to be indicated by $\bar{z}_{j}$ for $j^{th}$ individual $(j = 1,2,3,...,n)$. Independence between censoring scheme and life times of individuals has been presumed. Likelihood function can be described for hazard function with the involvement of univariate life time random variable of the $j^{th}$ individual as,

\[
\mathcal{L}_{j}(z_{j}) = \left\{ \begin{array}{ll}
f_{1}(z_{j}),  & \ ; \ z_{j} < \bar{z}_{j},\\
f_{2}(\bar{z}_{j}),    & \ ; \ z_{j} > \bar{z}_{j}.
\end{array}
\right.
\]
and likelihood function will be,
\begin{eqnarray}
\mathcal{L}(\underline{\Theta}, \underline{\beta_{0}}, \eta, \varepsilon) & = & \prod^{n_{1}}_{j = 1}f_{1}(z_{j}) \prod^{n_{2}}_{j = 1}f_{2}(\bar{z}_{j})
\end{eqnarray}
where,$\underline{\Theta}$, $\underline{\beta_{0}}$, $\eta$ and $\varepsilon$ are vector of baseline parameters and the vector of regression coefficients and frailty parameters respectively.
For $z_{j} < \bar{z}_{j}$ and $z_{j} > \bar{z}_{j}$, number of individuals corresponding to failure time $z_{j}$ will be $n_1$, $n_2$ respectively and the likelihood with the involvement of $j^{th}$ individual for the hazard rate,
\begin{eqnarray}
f_{1}(z_{j}) & = & -\frac{\partial \Lambda(z_{j})}{\partial z_{j}}  \notag
\end{eqnarray}
\begin{eqnarray}
f_{2}(\bar{z}_{j})  &=&  \Lambda(\bar{z}_{j}) \notag
\end{eqnarray}
substituting survival function $\Lambda(t_{j})$, hazard function $\phi_{0}(t_{j})$ for hazard rate setup (6.1) and by differentiating we get the likelihood function. We can compute the likelihood function for  IG-GW and GL-GW Models respectively.\\

\section{\large Bayesian Paradigm Using MCMC Techniques}
To achieve the estimates of model parameters, maximum likelihood estimates (MLEs) method having crucial importance. Inappropriately, both proposed models have high dimensional optimization problem. For that reason, MLEs method could not be a good choice. Consequently, we must employ Bayesian paradigm that has been earlier utilized by quite a lot of practitioners (see Ibrahim et al. (2001); Santos and Achcar (2010)). The joint posterior density function of parameters for given failure times is obtained as,
\begin{eqnarray}
\pi(\underline{\Theta}, \underline{\beta_{0}}, \eta, \varepsilon) \propto \mathcal{L}(\underline{\Theta}, \underline{\beta_{0}}, \eta, \varepsilon)\times  \nonumber \\ g_1(\zeta)g_2(\delta)g_3(\xi)g_4(\eta)g_5(\varepsilon) \prod_{i=1}^{m}p_{i}(\beta_{0 i\times 1})\nonumber
\end{eqnarray}
where $g_i(.)$ indicates the prior density function with known hyper parameters of corresponding argument for baseline parameters and frailty variance; $p_i(.)$ is prior density function for regression coefficient $\beta_{0i}$ and likelihood function is $\mathcal{L}(.)$. An important assumption here is, all the parameters are independently distributed. In similar way, joint posterior density function can be written for without frailty models. To estimate the parameters of the models, Metropolis-Hastings algorithms and Gibbs samplers have been used. Geweke test and Gelman-Rubin statistics have been used to monitored the convergence of a Markov chain to a stationary distribution.\\
Due to high-dimensions of conditional distributions, it is not unproblematic to integrate out. Thus, it has been considered that full conditional distributions can be obtained as, they are proportional to joint distribution of the parameter of the model. The conditional distribution for parameter $\zeta$ with frailty as,
\begin{eqnarray}
\pi_{1}(\zeta \mid \underline{\Theta}, \underline{\beta_{0}}, \eta, \varepsilon) & \propto & \mathcal{L}(\underline{\Theta}, \underline{\beta_{0}}, \eta, \varepsilon) \cdot g_{1}(\zeta)
\end{eqnarray}
similarly the conditional distributions for other parameters can be obtained.

\section{\large Numerical Studies}
\subsection{\large Monte Carlo Simulations}
A simulation study was executed to appraise the Bayesian estimation paradigm based on the generation of 25 artificial data from both models. In real life circumstances, there can be either one covariate or two covariates or many more exists. Therefore, single covariate $\underline{K}_{01}$ has been considered, follows normal distribution. The frailty variable $W$ is assumed to follow inverse Gaussian distribution for IG-GW model, and follow generalized Lindley distribution for GL-GW model. Independence between lifetimes of individuals has been considered. Samples are generated by using the subsequent mechanism,
\begin{enumerate}
	\item For IG-GW model, generate frailty variable $W$ form Inverse Gaussian distribution.
	\item To generate the values of frailty variable $W$ from generalized Lindley distribution for GL-GW model, we follow ensuing mechanism
	\begin{itemize}
		\item Generate $U_{j} \sim uniform(0, 1); j = 1, 2,..., 25$
		\item Generate $V_{j} \sim Gamma(25,\eta,\eta); j = 1, 2,...,25$
		\item Generate $Z_{j} \sim Gamma(25,\varepsilon,\varepsilon); j = 1, 2,...,25$
		\item If $U_{j}< \frac{\eta}{(\eta+\varepsilon)}$, then set $W_j = V_j$. Otherwise, set $W_j = Z_j$.
	\end{itemize}
	\item Generate $\underline{K}_{01}\sim B(25,0.6)$.
	\item For known covariates, compute $\rho = e^{\underline{K}_{01}\underline{\beta}_{01}}$.
	\item Lifetimes supposed to follows generalized Weibull baseline distribution for given frailty $W$. 25 values of lifetimes have been generated after using ensuing manners.\\
	Conditional survival function for lifetime $z_{j}$ $(j= 1, 2,..., n)$ for given frailty $W_{j}=w_{j}$ and covariate $\underline{K}_{01}$ is,\\
	\begin{eqnarray}
	S(z_{j}\mid w_{j}, \underline{K}_{01})&=& e^{-w_{j} \Phi_{0}(z_{j}) \rho}  \notag
	\end{eqnarray}
	equating $ S(z_{j}\mid w_{j}, \underline{K}_{01})$ to random number, say $v_{j}\sim U(0,1)$ over $z_{j}\in{\rm I\!R^{+}}$ we get,
\begin{eqnarray}
z_{j}&=& \left( -\frac{1}{\delta}\log(1-(1-v_{j}^{\frac{1}{w_{j}\rho_{j}}})^{\frac{1}{\zeta}})\right)^{\frac{1}{\xi}}  \notag
\end{eqnarray}	\item Censoring time $\bar{z}_{j}\sim exp(0.1); j = 1, 2,...,25 $ for both models.
	\item Observe the $j^{th}$ survival time $z^{*}_{j}=min(z_{j},\bar{z}_{j})$ and the  censoring indicator $\delta_{j}$ for the $j^{th}$ individual $(j=1,2,...,25)$ where, \\
	\[
	\delta_{j} = \left\{ \begin{array}{ll}
	1, & ; \ z_{j} < \bar{z}_{j}\\
	0, & ; \ z_{j} > \bar{z}_{j}
	\end{array}
	\right.
	\]
	thus we have data consisting of 25 pairs of survival times $z_{j}^{*}  $ and the censoring indicator $\delta_{j}$.
\end{enumerate}
Concurrently, with different priors and starting points, two chains have been operated.  Both chains recapitulated 100,000 times. Gelman-Rubin test values are very close to one. Due to small values of Geweke test statistic and corresponding p-values the chains reach stationary distribution for both prior sets. The estimates of parameters were the same for both the priors, no impact of prior distributions has been found on posterior summaries. Here, the analysis for one chain has been exhibited because both the chains have shown the same results. Tables 1 and 2 present the estimates, standard error and the credible intervals of the parameters for the IG-GW, GL-GW Models based on the simulation study. The Gelman-Rubin convergence statistic values are nearly equal to one and also the Geweke test values are quite small, and the corresponding p-values are large enough to say that the chain attains stationary distribution.

\section{Application}
We exemplify the potentiality of the proposed models with three real-world datasets via Bayesian estimation paradigm. The parameter estimation was performed by concerning hybrid Metropolis-Hastings (M-H) algorithm based on normal transition kernels. Vague priors have been used. For frailty parameters, gamma distribution with very small shape and scale parameters (say, 0.0001) has been used. Additionally, it can be considered, regression coefficients are normally distributed with mean zero and high variance (say 1000). Because of no information about baseline parameter, the prior distribution corresponding to baseline parameters are also considered flat. We considered two different vague prior distributions for baseline parameters, one is gamma distribution with shape and scale hyperparameters $\epsilon_1, \epsilon_2$ respectively and another is uniform distribution with interval $(\nu_1,\nu_2)$. All the hyperparameters are known. Under the Bayesian paradigm, for both models, two parallel chains have been run. Also, two sets of prior distributions have been used with different starting points. It can be said that estimates are independent of the different prior distributions because, for both sets of priors, estimates of parameters are approximately similar. We got almost similar convergence rate of Gibbs sampler for both sets of priors. Here, the analysis for one chain has been exhibited because both the chains have shown the same results.
\subsection{Bladder Cancer Data}
   First dataset that we have considered, is bladder cancer (Andrews and Hertzberg (1985)); given in R-package survival, named as bladder1. This data consists of 292 patients. Remaining Details of data given as below:
   \begin{itemize}
   	\item Survival time $=$ stop $-$ start
   	\item Censoring indicator: 1=uncensored, 0=censored (0 and 3). Remove number 2 completely from the data
   	\item Covariates:\\
   	$K_{01}$ : if Pyridoxine treatment Present, then $K_{01}=1$, and 0, otherwise\\
   	$K_{02}$ : if Thiotepa treatment present, then $K_{03}= 1$, and 0, otherwise\\
   	$K_{03}$ : Number of tumors\\
   	$K_{04}$ : Size of the largest initial tumor\\
   	$K_{05}$ : Total number of recurrences
   \end{itemize}
Here, the Kolmogorov-Smirnov (K-S) test has been applied to check the goodness of fit of the data set for baseline distributions. Consequently, on the basis of p-values of K-S test, given in Table 3 and  it is clear that there is no statistical evidence to reject the hypothesis that data are from the IG-GW and GL-GW Models. Figure 1 shows the parametric plot with semi-parametric plot for IG-GW and GL-GW Models with frailty and both lines are close to each other. Tables 6-7 contained the values of posterior mean and the standard error with 95$\%$ credible intervals, the Gelman-Rubin statistics values and the Geweke test with p-values for IG-GW and GL-GW Models. The Gelman-Rubin convergence statistic values are closely equal to one. The Geweke test statistic values are somewhat small, and the corresponding p-values are large enough to say that the chains reach stationary distribution. For IG-GW Model the credible interval of all regression coefficients do not contain zero. It indicates that all covariates have a significant effect on IG-GW model.  With negative value, it is being indicated that pyridoxine treatment, thiotepa treatment, and total number of tumors are significant factors for bladder cancer, having negative effects. But, with positive value size of the largest initial tumor and total number of recurrences  have a positive significant effect with a higher chance of cancer. For GL-GW Model, the credible interval of $\beta_{01}$, $\beta_{02}$, and $\beta_{05}$ regression coefficients do not contain zero. Consequently, pyridoxine, thiotepa treatments, and total number of recurrences are the Noteworthy  factors. With negative value, it is being indicated that pyridoxine and thiotepa treatments are significant factors for bladder cancer, having negative effects. With positive value total number of recurrences  has a positive significant effect with a higher chance of cancer. The AIC, BIC, AICc, and HQIC  values, given in Table 12, have been used to compare both models. IG-GW Model holds the lowest possible values of AIC, BIC, AICc, and HQIC.

\subsection{Lung Cancer Data}
Second dataset that we have considered, is lung cancer (Loprinzi et al. (1994)); given in R-package survival, named as lung. This data consists of 226 patients. Remaining Details of data given as below:
\begin{itemize}
	\item Survival time: time
	\item Censoring indicator: status
    \item Covariates:\\
    $K_{01}$ : Age\\
	$K_{02}$ : Sex (Female-0, Male-1)\\
	$K_{03}$ : ECOG performance score rated by physician (0-5: good-0,\dots,dead-5)\\
	$K_{04}$ : Karnofsky performance score rated by physician (0-100: bad-0,\dots,good-100)
\end{itemize}
Similarly, as bladder cancer data analysis, here, the K-S test has been applied to check the goodness of fit. Consequently, on the basis of p-values of K-S test, given in Table 4 and Figure 2, it is clear that there is no statistical evidence to reject the hypothesis that data are from the IG-GW and GL-GW Models.
 Tables 8-9 contained the values of posterior mean and the standard error with 95$\%$ credible intervals, the Gelman-Rubin statistics values and the Geweke test with p-values for IG-GW and GL-GW Models. The Gelman-Rubin convergence statistic values are closely equal to one. The Geweke test statistic values are somewhat small, and the corresponding p-values are large enough to say that the chains reach stationary distribution.
  For IG-GW and GL-GW Models, the credible interval of $\beta_{01}$, $\beta_{03}$, and $\beta_{04}$ regression coefficients do not contain zero. It indicates that age, ECOG performance score, and karnofsky performance score have a noteworthy effect on both models. Consequently, for IG-GW Model, with positive value of the regression coefficient of age, ECOG performance score, and karnofsky performance score have a positive significant effect with a higher chance of cancer. While for GL-GW Model, With negative value, it is being indicated that age is noteworthy factors for lung cancer, having negative effects. With positive value of the regression coefficient of ECOG performance score, and karnofsky performance score have a positive significant effect with a higher chance of cancer. The AIC, BIC, AICc, and HQIC  values, given in Table 13, have been used to compare both models. IG-GW Model holds the lowest possible values of AIC, BIC, AICc, and HQIC.

\subsection{Ovarian Cancer Data}
For illustrating the real-life application of both proposed frailty models, we consider third data, is ovarian cancer dataset (Edmonson, et al (1979)). This data consists of 26 patients. Remaining Details of data given as below:
\begin{itemize}
	\item Survival time: futime
	\item Censoring indicator: fustat
	\item Covariates:\\
	$K_{01}$  : Age \\
	$K_{02}$  : if Treatment 2, then $K_{02} =1$, and 0, otherwise\\
	$K_{03}$  : if Residual disease code is 2, $K_{03} = 1$, and 0, otherwise \\
	$K_{04}$  : ECOG performance status (better-1, otherwise-0)
	\end{itemize}
Similarly, as bladder and lung cancers data analysis, here, the K-S test has been applied to check the goodness of fit. Consequently, on the basis of p-values of K-S test, given in Table 5 and Figure 3, it is clear that there is no statistical evidence to reject the hypothesis that data are from the IG-GW and GL-GW Models.
Tables 10-11 contained the values of posterior mean and the standard error with 95$\%$ credible intervals, the Gelman-Rubin statistics values and the Geweke test with p-values for IG-GW and GL-GW Models. The Gelman-Rubin convergence statistic values are closely equal to one. The Geweke test statistic values are somewhat small, and the corresponding p-values are large enough to say that the chains reach stationary distribution.
For IG-GW and GL-GW Models, the credible interval of $\beta_{02}$, $\beta_{03}$, and $\beta_{04}$ regression coefficients contain zero. It indicates that treatment, residual disease, and ECOG performance have no noteworthy effect on both models. Consequently, for IG-GW Model, with positive value of the regression coefficient of age has a positive significant effect with a higher chance of cancer. The AIC, BIC, AICc, and HQIC  values, given in Table 14, have been used to compare both models. IG-GW Model holds the lowest possible values of AIC, BIC, AICc, and HQIC.

\subsection{\textcolor{red}{HIV-infected Patients Data}}
\textcolor{red}{
In the final data set, HIV-infected individuals are used to demonstrating the real-world use of the suggested frailty models. Manyau et al.(2020) published a retrospective review of medical records of adult HIV-infected patients treated in Zimbabwe for high-grade large cell non-Hodgkin's lymphoma between 2015-–2017. We look at a sub-sample of 91 individuals who had three or more treatment cycles. The variable of interest is overall survival (OS), which is measured in months from the time of lymphoma diagnosis to death. Twenty-six individuals were right-censored 20 months following diagnosis, as planned in the research. Sex and age are included as observed covariates in this study.} 

\section{\large Concluding Remarks}
The motivation here was that the frailty model might be more suitable to account for heterogeneity and natural interpretation. Because of this motivation, we proposed frailty based models named as IG-GW Model and GL-GW Model. We assumed IG and GL distributions for frailty and GW as baseline distribution. The methodology was implemented by using hybrid M-H algorithm via hand-written program in R-software to achieve the estimates of models. The estimates of frailty parameters are high in both models for all three datasets. The variances of IG-GW and GL-GW models for bladder cancer data are 	10.98781 and 10.20371, for lung cancer data are 1.13022 and 23.94934, and for ovarian cancer data are 2.34760 and 14.52641 correspondingly. This indicates that there is a strong evidence that heterogeneity is present among the study population. To take the decision about all models, different tools have been utilized. With the lowest value of AIC, BIC, AICc, and HQIC from Tables 12,13, and 14 it can be said that IG-GW Model is better than the GL-GW Model to such kind of datasets.

\section*{Appendix: Tables and Figures}

\begin{table}[h]
  {\fontsize{10}{10}\selectfont
  \caption{Posterior Summary of Inverse Gaussian Frailty with Baseline Generalized Weibull  (Simulation Study: IG-GW Model)}
     	
   \centering
  \begin{tabular}{|c  |r |r |r |r |r|r|r|}
    \hline
   $ $ & $ $ & $ $ & $ $ & $ $ & $ $ & $ $ &  $ $ \ \\
   $Parameter$& $Estimate$ & $s.e.$ & $L.C.L.$ &$U.C.L.$&$Geweke$& $p-value$&$Gelman$ \ \\
   $ $ & $ $ & $ $ & $ $ & $ $ & $ test$ & $ $ &  $ Rubin \ test$ \ \\  % inserts table
     			
   	\hline
    \multicolumn{2}{|l}{burn \ in \ period = 6900;} & \multicolumn{6}{l|}{autocorrelation \ lag = 400}\\
    \hline
     			
   	$\zeta$&	2.13412&	0.24965&	1.58873&	2.48424&	-0.00748&	0.49701&	1.00000\\
   	$\delta$&	0.70744&	0.05312&	0.60868&	0.79588&	-0.00270&	0.49892&	1.00008\\
   	$\xi$&	1.53737&	0.10334&	1.32764&	1.68748&	0.00241&	0.50096&	1.01757\\
   	$\eta$&	14.79766&	1.56816&	12.15943&	17.61182&	0.01332&	0.50531&	1.00109\\
   	$\beta_{01}$&	2.74224&	0.52037&	1.64587&	3.70262&	0.00073&	0.50029&	1.00037\\[1ex]
     			\hline
     	\end{tabular}}
     	\label{table:nonlin}
     \end{table}

\begin{table}[h]
	{\fontsize{10}{10}\selectfont
		\caption{Posterior Summary of Generalized Lindley Frailty with Baseline Generalized Weibull  (Simulation Study: GL-GW Model)}
		
		\centering
		\begin{tabular}{|c  |r |r |r |r |r|r|r|}
			\hline
			$ $ & $ $ & $ $ & $ $ & $ $ & $ $ & $ $ &  $ $ \ \\
			$Parameter$& $Estimate$ & $s.e.$ & $L.C.L.$ &$U.C.L.$&$Geweke$& $p-value$&$Gelman$ \ \\
			$ $ & $ $ & $ $ & $ $ & $ $ & $ test$ & $ $ &  $ Rubin \ test$ \ \\  % inserts table
			
			\hline
			\multicolumn{2}{|l}{burn \ in \ period = 6900;} & \multicolumn{6}{l|}{autocorrelation \ lag = 400}\\
			\hline
			
$\zeta$&	2.13555&	0.37240&	1.53740&	3.00573&	-0.00961&	0.49617&	1.00023\\
$\delta$&	0.90654&	0.05568&	0.80662&	0.99519&	-0.01866&	0.49256&	1.01960\\
$\xi$&	1.50179&	0.10389&	1.30911&	1.68844&	0.01700&	0.50678&	1.00010\\
$\eta$&	3.85160&	0.54070&	3.02752&	4.87382&	0.01135&	0.50453&	0.99996\\
$\varepsilon$&	1.16993&	0.10439&	1.00863&	1.37432&	0.00647&	0.50258&	0.99998\\
$\beta_{01}$&	-0.07141&	0.16341&	-0.37496&	0.18906&	-0.00615&	0.49755&	0.99996\\[1ex]
			\hline
	\end{tabular}}
	\label{table:nonlin}
\end{table}

\begin{table}[ht]
	\caption{p-value of K-S statistics for goodness of fit\\
		test for Bladder Cancer data set}
	% title of Table
	\centering  % used for centering table
	\begin{tabular}{|c  |r |} % centered columns (4 columns)
		\hline                       %inserts double horizontal lines
		Model & p-value  \\ [0.5ex] % inserts table
		%heading
		\hline                  % inserts single horizontal line
		$IG-GW Model$   & 0.9634    \\ % inserting body of the table
		$GL-GW Model$   & 1.0000 \\[1ex]    % [1ex] adds vertical space
		\hline %inserts single line
	\end{tabular}
	\label{table:nonlin} % is used to refer this table in the text
\end{table}

\begin{table}[ht]
	\caption{p-value of K-S statistics for goodness of fit\\
		test for Lung Cancer data set}
	% title of Table
	\centering  % used for centering table
	\begin{tabular}{|c  |r |} % centered columns (4 columns)
		\hline                       %inserts double horizontal lines
		Model & p-value  \\ [0.5ex] % inserts table
		%heading
		\hline                  % inserts single horizontal line
		$IG-GW Model$   & 0.4718    \\ % inserting body of the table
		$GL-GW Model$   & 0.3843 \\[1ex]    % [1ex] adds vertical space
		\hline %inserts single line
	\end{tabular}
	\label{table:nonlin} % is used to refer this table in the text
\end{table}

\begin{table}[ht]
	\caption{p-value of K-S statistics for goodness of fit\\
		test for Ovarian Cancer data set}
	% title of Table
	\centering  % used for centering table
	\begin{tabular}{|c  |r |} % centered columns (4 columns)
		\hline                       %inserts double horizontal lines
		Model &  p-value  \\ [0.5ex] % inserts table
		%heading
		\hline                  % inserts single horizontal line
		$IG-GW Model$   & 0.9985    \\ % inserting body of the table
		$GL-GW Model$   & 0.8690 \\[1ex]    % [1ex] adds vertical space
		\hline %inserts single line
	\end{tabular}
	\label{table:nonlin} % is used to refer this table in the text
\end{table}

\begin{table}[ht]
	\caption{\textcolor{red}{ p-value of K-S statistics for goodness of fit\\
		test for HIV data set}}
	% title of Table
	\centering  % used for centering table
	\begin{tabular}{|c  |r |} % centered columns (4 columns)
		\hline                       %inserts double horizontal lines
		Model &  p-value  \\ [0.5ex] % inserts table
		%heading
		\hline                  % inserts single horizontal line
		$IG-GW Model$   &  5.223e-07\\ % inserting body of the table
		$GL-GW Model$   &  2.089e-06\\[1ex]    % [1ex] adds vertical space
		\hline %inserts single line
	\end{tabular}
	\label{table:nonlin} % is used to refer this table in the text
\end{table}

\begin{table}[h]
	{\fontsize{10}{10}\selectfont
		\caption{Posterior Summary of Inverse Gaussian Frailty with Baseline Generalized Weibull for Bladder Cancer Data ( IG-GW Model)}
		
		\centering
		\begin{tabular}{|c  |r |r |r |r |r|r|r|}
			\hline
			$ $ & $ $ & $ $ & $ $ & $ $ & $ $ & $ $ &  $ $ \ \\
			$Parameter$& $Estimate$ & $s.e.$ & $L.C.L.$ &$U.C.L.$&$Geweke$& $p-value$&$Gelman$ \ \\
			$ $ & $ $ & $ $ & $ $ & $ $ & $ test$ & $ $ &  $ Rubin \ test$ \ \\  % inserts table
			
			\hline
			\multicolumn{2}{|l}{burn \ in \ period = 6900;} & \multicolumn{6}{l|}{autocorrelation \ lag = 400}\\
			\hline
			
$\zeta$&	4.88362&	0.09419&	4.69987&	5.06670&	0.00032&	0.50013&	1.00467\\
$\delta$&	0.40587&	0.00969&	0.38776&	0.42640&	-0.01456&	0.49419&	1.00493\\
$\xi$&	1.08035&	0.02001&	1.02969&	1.10737&	0.00737&	0.50294&	1.01531\\
$\eta$&	10.98781&	0.33545&	10.35480&	11.70782&	-0.01291&	0.49485&	1.02830\\
$\beta_{01}$&	-0.94788&	0.09088&	-1.17363&	-0.81934&	-0.00834&	0.49667&	1.02763\\
$\beta_{02}$&	-32.26824&	2.57038&	-36.91468&	-27.81438&	-0.00749&	0.49667&	1.00948\\
$\beta_{03}$&	-0.04680&	0.01013&	-0.06735&	-0.03106&	-0.00924&	0.49631&	1.03978\\
$\beta_{04}$&	0.33631&	0.03933&	0.24236&	0.39505&	0.00551&	0.50220&	1.00463\\
$\beta_{05}$&	0.11057&	0.01343&	0.07886&	0.12920&	-0.00758&	0.49698&	1.00495\\[1ex]
			\hline
	\end{tabular}}
	\label{table:nonlin}
\end{table}
\begin{table}[h]
	{\fontsize{10}{10}\selectfont
		\caption{Posterior Summary of Generalized Lindley Frailty with Baseline Generalized Weibull for Bladder Cancer Data ( GL-GW Model)}
		
		\centering
		\begin{tabular}{|c  |r |r |r |r |r|r|r|}
			\hline
			$ $ & $ $ & $ $ & $ $ & $ $ & $ $ & $ $ &  $ $ \ \\
			$Parameter$& $Estimate$ & $s.e.$ & $L.C.L.$ &$U.C.L.$&$Geweke$& $p-value$&$Gelman$ \ \\
			$ $ & $ $ & $ $ & $ $ & $ $ & $ test$ & $ $ &  $ Rubin \ test$ \ \\  % inserts table
			
			\hline
			\multicolumn{2}{|l}{burn \ in \ period = 6900;} & \multicolumn{6}{l|}{autocorrelation \ lag = 400}\\
			\hline
			
$\zeta$&	3.59945&	0.52112&	2.69821&	4.79571&	-0.00161&	0.49936&	1.01513\\
$\delta$&	0.36178&	0.05143&	0.27590&	0.46073&	-0.00042&	0.49983&	1.03187\\
$\xi$&	1.03683&	0.06312&	0.89617&	1.14380&	-0.00170&	0.49932&	1.00050\\
$\eta$&	11.81844&	0.66821&	10.46562&	13.05950&	0.00526&	0.50210&	1.00039\\
$\varepsilon$&	0.90364&	0.07756&	0.80332&	1.08054&	-0.00114&	0.49954&	1.02204\\
$\beta_{01}$&	-0.81633&	0.09315&	-1.02217&	-0.67780&	-0.00563&	0.49776&	1.00252\\
$\beta_{02}$&	-32.85344&	2.51813&	-37.41031&	-28.37949&	-0.00364&	0.49776&	1.00052\\
$\beta_{03}$&	-0.01168&	0.07083&	-0.14097&	0.14039&	0.00452&	0.50180&	1.00174\\
$\beta_{04}$&	0.09301&	0.06995&	-0.04830&	0.25852&	-0.00131&	0.49948&	1.00169\\
$\beta_{05}$&	0.03820&	0.00957&	0.02182&	0.05749&	-0.00929&	0.49629&	1.00167\\[1ex]
			\hline
	\end{tabular}}
	\label{table:nonlin}
\end{table}

\begin{table}[h]
	{\fontsize{10}{10}\selectfont
		\caption{Posterior Summary of Inverse Gaussian Frailty with Baseline Generalized Weibull for Lung Cancer Data ( IG-GW Model)}
		
		\centering
		\begin{tabular}{|c  |r |r |r |r |r|r|r|}
			\hline
			$ $ & $ $ & $ $ & $ $ & $ $ & $ $ & $ $ &  $ $ \ \\
			$Parameter$& $Estimate$ & $s.e.$ & $L.C.L.$ &$U.C.L.$&$Geweke$& $p-value$&$Gelman$ \ \\
			$ $ & $ $ & $ $ & $ $ & $ $ & $ test$ & $ $ &  $ Rubin \ test$ \ \\  % inserts table
			
			\hline
			\multicolumn{2}{|l}{burn \ in \ period = 6900;} & \multicolumn{6}{l|}{autocorrelation \ lag = 400}\\
			\hline
			
$\zeta$&	5.53026&	0.23185&	5.08251&	5.94977&	0.00320&	0.50128&	1.02517\\
$\delta$&	0.07740&	0.00322&	0.07129&	0.08366&	0.00659&	0.50263&	1.00230\\
$\xi$&	0.39925&	0.01152&	0.37761&	0.42088&	0.00969&	0.50386&	1.00399\\
$\eta$&	1.13022&	0.16785&	0.81815&	1.46123&	0.00541&	0.50216&	1.00217\\
$\beta_{01}$&	0.00505&	0.00106&	0.00309&	0.00674&	-0.00724&	0.49711&	1.00232\\
$\beta_{02}$&	0.17451&	0.19442&	-0.21630&	0.57903&	-0.02038&	0.49711&	1.00084\\
$\beta_{03}$&	0.71322&	0.14146&	0.41311&	0.98141&	-0.00841&	0.49664&	1.00359\\
$\beta_{04}$&	0.02884&	0.00312&	0.02358&	0.03567&	-0.00030&	0.49988&	1.01064\\[1ex]
			\hline
	\end{tabular}}
	\label{table:nonlin}
\end{table}
\begin{table}[h]
	{\fontsize{10}{10}\selectfont
		\caption{Posterior Summary of Generalized Lindley Frailty with Baseline Generalized Weibull for Lung Cancer Data ( GL-GW Model)}
		
		\centering
		\begin{tabular}{|c  |r |r |r |r |r|r|r|}
			\hline
			$ $ & $ $ & $ $ & $ $ & $ $ & $ $ & $ $ &  $ $ \ \\
			$Parameter$& $Estimate$ & $s.e.$ & $L.C.L.$ &$U.C.L.$&$Geweke$& $p-value$&$Gelman$ \ \\
			$ $ & $ $ & $ $ & $ $ & $ $ & $ test$ & $ $ &  $ Rubin \ test$ \ \\  % inserts table
			
			\hline
			\multicolumn{2}{|l}{burn \ in \ period = 6900;} & \multicolumn{6}{l|}{autocorrelation \ lag = 400}\\
			\hline
			
$\zeta$&	6.48069&	0.26746&	5.97924&	7.04540&	-0.00445&	0.49822&	1.00245\\
$\delta$&	0.13856&	0.00658&	0.12648&	0.15176&	-0.00143&	0.49943&	1.00870\\
$\xi$&	0.34743&	0.01062&	0.32507&	0.36873&	0.01191&	0.50475&	1.00496\\
$\eta$&	24.43131&	1.36498&	21.77210&	27.11021&	0.00949& 0.50378&	1.02267\\
$\varepsilon$&	0.24462&	0.02546&	0.20268&	0.29325&	-0.00463&	0.49815&	1.00940\\
$\beta_{01}$&	-0.00478&	0.00108&	-0.00690&	-0.00311&	0.00116&	0.50046&	1.00908\\
$\beta_{02}$&	0.10527&	0.17279&	-0.21961&	0.45616&	-0.00634&	0.50046&	1.00165\\
$\beta_{03}$&	0.62652&	0.11803&	0.39099&	0.85356&	0.00196&	0.50078&	1.00585\\
$\beta_{04}$&	0.02780&	0.00236&	0.02426&	0.03256&	-0.01391&	0.49445&	1.01666\\[1ex]
			\hline
	\end{tabular}}
	\label{table:nonlin}
\end{table}

 \begin{table}[h]
 	{\fontsize{10}{10}\selectfont
 		\caption{Posterior Summary of Inverse Gaussian Frailty with Baseline Generalized Weibull for Ovarian Cancer Data ( IG-GW Model)}
 		
 		\centering
 		\begin{tabular}{|c  |r |r |r |r |r|r|r|}
 			\hline
 			$ $ & $ $ & $ $ & $ $ & $ $ & $ $ & $ $ &  $ $ \ \\
 			$Parameter$& $Estimate$ & $s.e.$ & $L.C.L.$ &$U.C.L.$&$Geweke$& $p-value$&$Gelman$ \ \\
 			$ $ & $ $ & $ $ & $ $ & $ $ & $ test$ & $ $ &  $ Rubin \ test$ \ \\  % inserts table
 			
 			\hline
 			\multicolumn{2}{|l}{burn \ in \ period = 6900;} & \multicolumn{6}{l|}{autocorrelation \ lag = 400}\\
 			\hline
 			
$\zeta$&	4.14478&	0.30676&	3.53875&	4.71456&	-0.01248&	0.49502&	1.01011\\
$\delta$&	0.03603&	0.00329&	0.02961&	0.04263&	0.00605&	0.50241&	1.00432\\
$\xi$&	0.46823&	0.03209&	0.41202&	0.52906&	-0.00368&	0.49853&	1.00313\\
$\eta$&	2.34760&	0.34109&	1.69198&	3.10065&	0.00359&	0.50143&	1.00508\\
$\beta_{01}$&	0.04996&	0.01325&	0.02473&	0.07703&	-0.00339&	0.49865&	1.00317\\
$\beta_{02}$&	0.29916&	0.25462&	-0.16702&	0.76803&	-0.00401&	0.49865&	1.00528\\
$\beta_{03}$&	-1.10667&	0.72437&	-2.68438&	0.22915&	0.00675&	0.50269&	1.00102\\
$\beta_{04}$&	0.03500&	0.16584&	-0.35957&	0.23979&	-0.00623&	0.49751&	1.01808\\[1ex]
 			\hline
 	\end{tabular}}
 	\label{table:nonlin}
 \end{table}

\begin{table}[h]
	{\fontsize{10}{10}\selectfont
		\caption{Posterior Summary of Generalized Lindley Frailty with Baseline Generalized Weibull for Ovarian Cancer Data ( GL-GW Model)}
		
		\centering
		\begin{tabular}{|c  |r |r |r |r |r|r|r|}
			\hline
			$ $ & $ $ & $ $ & $ $ & $ $ & $ $ & $ $ &  $ $ \ \\
			$Parameter$& $Estimate$ & $s.e.$ & $L.C.L.$ &$U.C.L.$&$Geweke$& $p-value$&$Gelman$ \ \\
			$ $ & $ $ & $ $ & $ $ & $ $ & $ test$ & $ $ &  $ Rubin \ test$ \ \\  % inserts table
			
			\hline
			\multicolumn{2}{|l}{burn \ in \ period = 6900;} & \multicolumn{6}{l|}{autocorrelation \ lag = 400}\\
			\hline
			
$\zeta$&	4.34227&	0.31597&	3.70355&	4.99849&	0.00495&	0.50197&	1.00473\\
$\delta$&	0.05920&	0.00610&	0.04803&	0.07131&	0.00781&	0.50311&	1.00538\\
$\xi$&	0.50554&	0.03083&	0.44983&	0.56716&	0.00495&	0.50197&	1.00379\\
$\eta$&	15.92932&	0.92686&	14.15717&	17.63069&	0.00281&	0.50112&	1.00710\\
$\varepsilon$&	0.75235&	0.19683&	0.39118&	1.12461&	-0.00533&	0.49787&	1.02719\\
$\beta_{01}$&	0.01771&	0.00874&	-0.00136&	0.03168&	-0.00726&	0.49710&	1.00562\\
$\beta_{02}$&	0.02757&	0.05286&	-0.07024&	0.12076&	0.00514&	0.49710&	1.00086\\
$\beta_{03}$&	-1.04447&	0.68560&	-2.49074&	0.39063&	0.00223&	0.50089&	1.00080\\
$\beta_{04}$&	-0.00810&	0.02038&	-0.04559&	0.02851&	0.00858&	0.50342&	1.00001\\[1ex]
			\hline
	\end{tabular}}
	\label{table:nonlin}
\end{table}

\begin{table}[h]
	{\fontsize{10}{10}\selectfont
		\caption{\textcolor{red}{Posterior Summary of Inverse Gaussian Frailty with Baseline Generalized Weibull for HIV Data ( IG-GW Model)}}
		
		\centering
		\begin{tabular}{|c  |r |r |r |r |r|r|r|}
			\hline
			$ $ & $ $ & $ $ & $ $ & $ $ & $ $ & $ $ &  $ $ \ \\
			$Parameter$& $Estimate$ & $s.e.$ & $L.C.L.$ &$U.C.L.$&$Geweke$& $p-value$&$Gelman$ \ \\
			$ $ & $ $ & $ $ & $ $ & $ $ & $ test$ & $ $ &  $ Rubin \ test$ \ \\  % inserts table
			
			\hline
			\multicolumn{2}{|l}{burn \ in \ period = 6900;} & \multicolumn{6}{l|}{autocorrelation \ lag = 400}\\
			\hline
			
$\zeta$&	29.19799&	2.40945&	25.54654&	34.11914&	0.01277&	0.50509&	1.00870\\
$\delta$&	1.10431&	0.02197&	1.06226&	1.15093&	0.00912&	0.50364&	1.00226\\
$\xi$&	0.50281&	0.01515&	0.47383&	0.53390&	0.00282&	0.50113&	1.02739\\
$\theta$&	29.63543&	2.30832&	25.66838&	33.81806&	0.00336&	0.50134&	1.00800\\
$\beta_{01}$&	0.87702&	0.13267&	0.66422&	1.12544&	-0.00668&	0.49734&	1.00019\\
$\beta_{02}$&	0.04006&	0.00263&	0.03556&	0.04450&	0.00181&	0.49734&	1.00351\\[1ex]
			\hline
	\end{tabular}}

	\label{table:nonlin}
\end{table}

\begin{table}[h]
	{\fontsize{10}{10}\selectfont
		\caption{\textcolor{red}{Posterior Summary of Generalized Lindley Frailty with Baseline Generalized Weibull for HIV Data ( GL-GW Model)}}
		
		\centering
		\begin{tabular}{|c  |r |r |r |r |r|r|r|}
			\hline
			$ $ & $ $ & $ $ & $ $ & $ $ & $ $ & $ $ &  $ $ \ \\
			$Parameter$& $Estimate$ & $s.e.$ & $L.C.L.$ &$U.C.L.$&$Geweke$& $p-value$&$Gelman$ \ \\
			$ $ & $ $ & $ $ & $ $ & $ $ & $ test$ & $ $ &  $ Rubin \ test$ \ \\  % inserts table
			
			\hline
			\multicolumn{2}{|l}{burn \ in \ period = 6900;} & \multicolumn{6}{l|}{autocorrelation \ lag = 400}\\
			\hline
			
$\zeta$&	4.04738&	0.67025&	2.73310&	5.41978&	-0.00689&	0.49725&	1.00439\\
$\delta$&	0.10705&	0.02575&	0.05826&	0.14826&	-0.00349&	0.49861&	1.00920\\
$\xi$&	0.95953&	0.05845&	0.84753&	1.07224&	-0.00815&	0.49675&	1.01081\\
$\theta$&	109.61520&	15.61260&	82.32069&	137.47070&	-0.01668&	0.49335&	1.00206\\
$\mu$&	2.63623&	0.43708&	1.83954&	3.49350&	-0.00912&	0.49636&	0.99998\\
$\beta_{01}$&	0.56543&	0.41571&	-0.23340&	1.37388&	-0.00287&	0.49885&	1.00060\\
$\beta_{02}$&	0.04847&	0.00964&	0.03164&	0.06797&	0.00393&	0.49885&	1.00004\\[1ex]
			\hline
	\end{tabular}}
	
	\label{table:nonlin}
\end{table}

\begin{table}[ht]
	\begin{center}
		\caption{AIC, BIC, AICc, and HQIC Comparison for\\
			 Bladder Cancer Data }
	\end{center}
	% title of Table
	\centering  % used for centering table
	\begin{tabular}{|c  |r |r |r |r|} % centered columns (4 columns)
		\hline                        %inserts double horizontal lines
		Model  & AIC   &   BIC   & AICc & HQIC \\ [0.5ex] % inserts table
		%heading
		\hline                  % inserts single horizontal line
		$IG-GW Model$ &		1193.204&	1226.295&	1193.842&  1206.459\\
		$GL-GW Model$ &		1199.637&	1236.404&	1200.419&  1214.364\\[1ex]
		% [1ex] adds vertical space
		\hline %inserts single line	
	\end{tabular}
	\label{table:nonlin} % is used to refer this table in the text
\end{table}

\begin{table}[ht]
	\begin{center}
		\caption{AIC, BIC, AICc, and HQIC Comparison for\\
			Lung Cancer Data }
	\end{center}
	% title of Table
	\centering  % used for centering table
	\begin{tabular}{|c  |r |r |r |r|} % centered columns (4 columns)
		\hline                        %inserts double horizontal lines
		Model  & AIC   &   BIC   & AICc & HQIC \\ [0.5ex] % inserts table
		%heading
		\hline                  % inserts single horizontal line
		$IG-GW Model$ &		2315.666&	2343.031&	2316.33&  2326.709\\
		$GL-GW Model$ &		2316.029&	2346.813&	2316.862&  2328.452\\[1ex]
		% [1ex] adds vertical space
		\hline %inserts single line	
	\end{tabular}
	\label{table:nonlin} % is used to refer this table in the text
\end{table}

\begin{table}[ht]
	\begin{center}
		\caption{AIC, BIC, AICc, and HQIC Comparison for\\
			Ovarian Cancer Data }
	\end{center}
	% title of Table
	\centering  % used for centering table
	\begin{tabular}{|c  |r |r |r |r|} % centered columns (4 columns)
		\hline                        %inserts double horizontal lines
		Model  & AIC   &   BIC   & AICc & HQIC \\ [0.5ex] % inserts table
		%heading
		\hline                  % inserts single horizontal line
		$IG-GW Model$ &		204.599&	214.665&	213.0704&  207.4981\\
		$GL-GW Model$ &		208.501&	219.824&	219.7514&  211.762\\[1ex]
		% [1ex] adds vertical space
		\hline %inserts single line	
	\end{tabular}
	\label{table:nonlin} % is used to refer this table in the text
\end{table}

\begin{table}[ht]
	\begin{center}
		\caption{\textcolor{red}{ AIC, BIC, AICc, and HQIC Comparison for\\
			HIV Data} }
	\end{center}
	% title of Table
	\centering  % used for centering table
	\begin{tabular}{|c  |r |r |r |r|} % centered columns (4 columns)
		\hline                        %inserts double horizontal lines
		Model  & AIC   &   BIC   & AICc & HQIC \\ [0.5ex] % inserts table
		%heading
		\hline                  % inserts single horizontal line
		$IG-GW Model$ &		504.6717&	519.7369&	505.6717&	510.7496\\
		$GL-GW Model$ &		503.1032&	520.6793&	504.4526&	510.1941\\[1ex]
		% [1ex] adds vertical space
		\hline %inserts single line	
	\end{tabular}
	\label{table:nonlin} % is used to refer this table in the text
\end{table}

 \newpage
   \begin{figure}[!htb]
   	\centering
   	\includegraphics[scale=0.5]{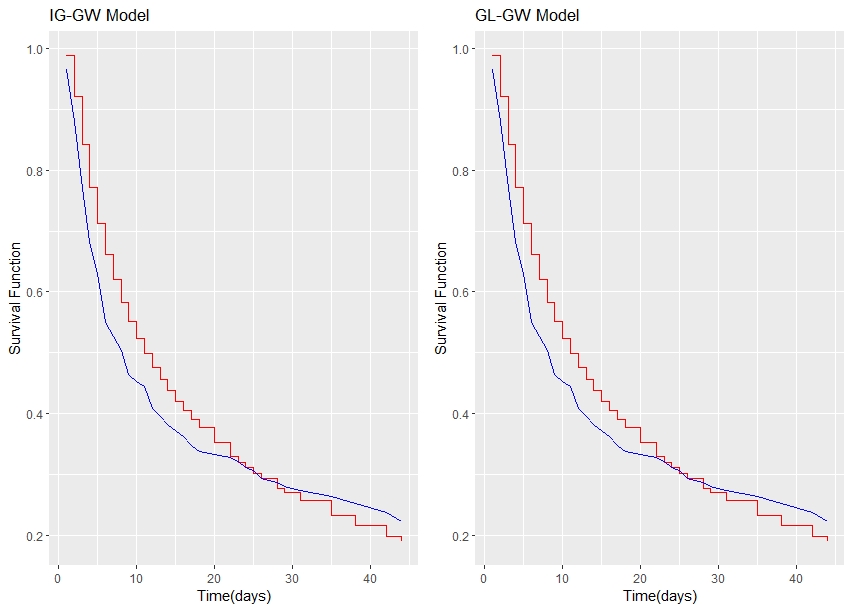}
   	\caption{K-S Plot for Bladder Cancer Data }
   	\label{fig:digraph}
   \end{figure}
   \begin{figure}[!htb]
	\centering
	\includegraphics[scale=0.5]{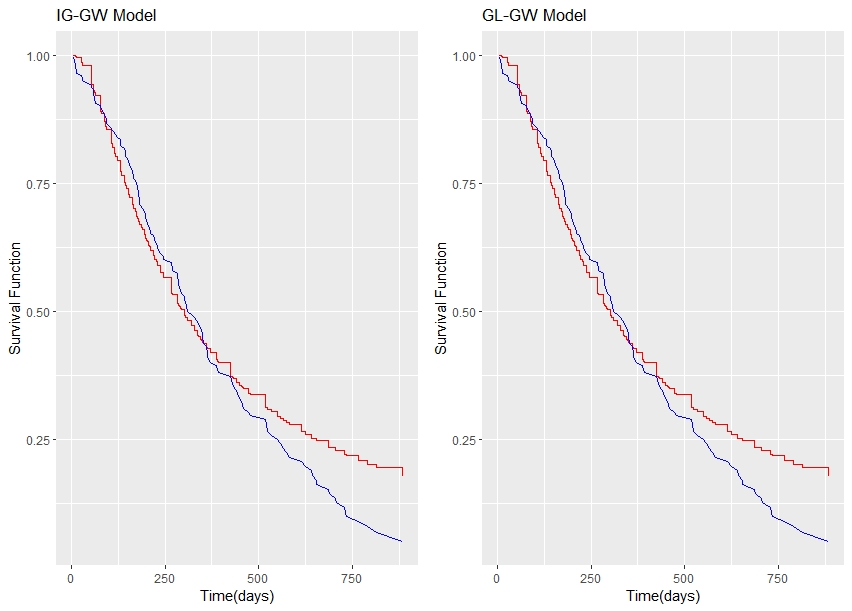}
	\caption{K-S Plot for Lung Cancer Data }
	\label{fig:digraph}
\end{figure}
   \begin{figure}[!htb]
	\centering
	\includegraphics[scale=0.5]{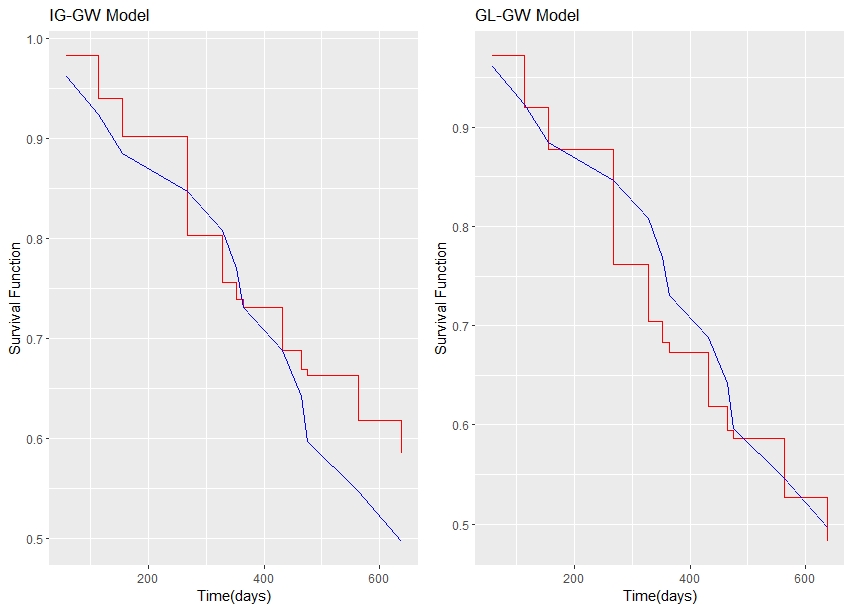}
	\caption{K-S Plot for Ovarian Cancer Data }
	\label{fig:digraph}
\end{figure}
\end{document}